\documentstyle[sprocl,epsf]{article}

\def\Journal#1#2#3#4{{#1} {\bf #2}, #3 (#4)}

\def\NPA{{\em Nucl. Phys.} A}
\def\NPB{{\em Nucl. Phys.} B}
\def\PLB{{\em Phys. Lett.}  B}
\def\PRL{\em Phys. Rev. Lett.}

\def\PRC{{\em Phys. Rev.} C}

\def\ZPA{{\em Z. Phys.} A}
\def\ANN{\em Ann. Phys.}
\def\FBS{\em Few-Body Sys. Suppl.}
\def\PR{\em Phys. Rep.}

\def\cpt{\chi -PT}
\def\lsc{\Lambda _\chi}

\def\be{\begin{equation}}
\def\ee{\end{equation}}
\def\bea{\begin{eqnarray}}
\def\eea{\end{eqnarray}}

\def\lsc{\Lambda _\chi}
\def\cpt{\chi PT}

\def\vkay{{\vec k}}
\def\vkayprime{{{\vec k}\, '}}
\def\vpee{{\vec p}}
\def\vpeeprime{{{\vec p}\, '}}

\begin{document}

\title{GAMMA-DEUTERON SCATTERING}

\author{S.R.~BEANE}

\address{Department of Physics, University of Maryland,\\
College Park, MD 20742-4111\\E-mail: sbeane@physics.umd.edu} 

\maketitle\abstracts{We discuss a recent computation of Compton
scattering on the deuteron at photon energies of order the pion
mass. An interaction kernel is computed in baryon chiral perturbation
theory and sewn to phenomenological deuteron wave functions. As an
appetizer, we consider a computation of the pion-deuteron scattering
length using this method.}

\section{Introduction}
\label{sec:intro}

\noindent 

\noindent Precise calculations of hadron processes are possible only
where a small dimensionless expansion parameter is identified. This is
the main motivation behind the ongoing intense effort to develop a
perturbative theory of nuclear interactions~\cite{monster}.  The
dimensionless parameters relevant to low energy QCD and therefore to
nuclear physics consist of ratios of external momenta to various
characteristic energy scales, like the nucleon mass.  Effective field
theory is the technology which develops a hierarchy of scales into a
perturbative expansion of physical observables. In a system with
broken symmetries this technology is especially powerful. When a
continuous symmetry is spontaneously broken there are massless
Goldstone modes which dominate the low-energy dynamics and couple only
derivatively. Hence at energies small relative to the characteristic
symmetry breaking scale, the interactions of the Goldstone bosons can
be ordered in an effective Lagrangian which is constrained by chiral
symmetry and in which each operator contains a nonvanishing number of
derivatives acting on the pion fields.  Observables computed from the
effective Lagrangian are therefore power series in momenta, with the
non-analyticities required by perturbative unitarity.

In this paper we describe several recent effective field theory
calculations whose objective is to extract nucleon properties from
nuclear scattering processes in a systematic way. We first discuss the
paradigmatic problem of the pion-deuteron scattering length and its
dependence on nucleon parameters.  We then describe a recent
calculation of Compton scattering on the deuteron at photon energies
of order the pion mass. Here the ultimate objective is to learn about
neutron polarizabilities from nuclear Compton scattering. The basic
power-counting scheme is reviewed in Sec.~\ref{sec:power}. In
Sec.~\ref{sec:app} we discuss the problem of pion-deuteron scattering
at threshold as a heuristic tool.  Sec.~\ref{sec:entre} is dedicated
to Compton scattering on the deuteron. We conclude in
Sec.~\ref{sec:conc}.

\section{Weinberg power-counting}
\label{sec:power}

\noindent At energies well below the chiral symmetry breaking scale,
$\lsc\sim 4 \pi f_\pi \sim {M} \sim {m_\rho}$, the interactions of
pions, photons and nucleons can be described systematically using an
effective field theory.  This effective field theory, known as chiral
perturbation theory ($\cpt \,$), reflects the observed QCD pattern of
symmetry breaking.  In QCD the chiral $SU(2)_L \times SU(2)_R$
symmetry is spontaneously broken.  Here we are interested in processes
where the typical momenta of all external particles is $p\ll\lsc$, so
we identify our expansion parameter as $p/\lsc$.  In QCD $SU(2)_L
\times SU(2)_R$ is softly broken by the small quark masses. This
explicit breaking implies that the pion has a small mass in the
low-energy theory.  Since ${m_\pi}/\lsc$ is then also a small
parameter, we have a dual expansion in $p/\lsc$ and ${m_\pi}/\lsc$. We
take $Q$ to represent either a small momentum {\it or} a pion mass.

In few-nucleon systems, a complication arises in $\cpt$ due to the
existence of shallow nuclear bound states and related infrared
singularities in $A$-nucleon reducible Feynman diagrams evaluated in
the static approximation~\cite{weinnp}. The fundamental problem is
that nuclear physics introduces a new mass scale, the nuclear binding
energy, which is very small compared to a typical hadronic scale like
$\lsc$.  One way to overcome this difficulty is to adopt a modified
power-counting scheme in which $\cpt$ is used to calculate an
effective potential which generally consists of all $A$-nucleon
irreducible graphs. The $S$-matrix, which includes all reducible
graphs as well, is then obtained through iteration by solving a
Lippmann-Schwinger equation~\cite{weinnp}. This version of nuclear
effective theory is known as the Weinberg formulation.  To date the
Weinberg formulation can be carried through rigorously only using
finite cutoff regularization~\cite{ordonez,vk,ray,lepage}.  This
limitation has spawned an intense theoretical effort geared at
formulating an effective field theory for low-lying bound states which
is verifiably (analytically) consistent in the sense of
renormalization~\cite{monster}. One result of this effort is a new
power-counting scheme in which all nonperturbative physics responsible
for the presence of low-lying bound states arises from the iteration
of a single operator in the effective theory, while all other effects,
including all higher dimensional operators {\it and} pion exchange,
are treated perturbatively~\cite{pc,ksw}. This version of the
effective theory is known as the Kaplan-Savage-Wise (KSW)
formulation. This is relevant here because Compton scattering on the
deuteron has been computed to next-to-leading order in the KSW
formulation~\cite{martinetal}.  We will discuss this result and its
relation to our calculation.  A comprehensive and up-to-date review of
nuclear applications of effective field theories can be found in
Ref.~\cite{monster}.

In the interactions of the deuteron with pionic and electromagnetic
probes, Weinberg power-counting is encoded in the following simple
diagrammatic rules:

\begin{itemize}
\item A nucleon propagator contributes $Q^{-1}$;

\item A pion propagator contributes $Q^{-2}$;

\item Each derivative or power of the pion mass at a vertex
contributes $Q$. Therefore, an operator insertion from the $\pi$-N
Lagrangian with $n$ derivatives or powers of $m_\pi$, ${\cal L}_{\pi
N}^{(n)}$, contributes $Q^n$;

\item Each loop integral contributes $Q^4$;

\item Each deuteron wavefunction, $\bigtriangleup$ or $\bigtriangledown$,
contributes $Q^{-{1\over 2}}$.
\end{itemize}

It should be noted that typical nucleon momenta inside the deuteron
are small---on the order of $\sqrt{MB}$ or $m_\pi$, with $B$ the
deuteron binding energy---and consequently, {\it a priori} we expect
no convergence problems in the $\cpt$ expansion of any low-momentum
electromagnetic or pionic probe of the deuteron.  Although in
principle we could use wavefunctions computed in $\cpt$, we will
consider wavefunctions generated using modern nucleon-nucleon
potentials. Generally we find that any wavefunction with the correct
binding energy gives equivalent results to within the theoretical
error expected from neglected higher orders in the chiral expansion.
Presumably we are insensitive to short distance components of the
wavefunction because we are working at low energies and the deuteron
is a large object.

Weinberg power-counting has led to fruitful computation of many pionic
and photonic probes of the two-nucleon system~\cite{monster},
including the pion-deuteron scattering length~\cite{wein1,silas1},
neutral pion photoproduction on the deuteron at
threshold~\cite{silas2}, Compton scattering on the
deuteron~\cite{silas3}, as well as $pn$ radiative capture~\cite{rho1}
and the solar burning process $p p \rightarrow d e^+ \nu$~\cite{rho2}.

\section{Appetizer: threshold $\pi$-deuteron scattering}
\label{sec:app}

\begin{figure}[t]
   \vspace{0.5cm} \epsfysize=3.5cm
   \centerline{\epsffile{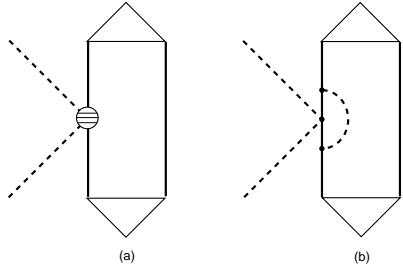}}
   \centerline{\parbox{11cm}{\caption{\label{fig1} Contributions to
   the $\pi$-deuteron scattering length at order $Q^2$ (a) and $Q^3$
   (b).  The dots are vertices from ${\cal L}_{\pi N}^{(1)}$ and the
   sliced blob is from ${\cal L}_{\pi N}^{(2)}$.  All topologies are
   not shown.}}}
\end{figure}

Effective field theory relates scattering processes involving a single
nucleon to nuclear scattering processes.  For instance, one can relate
$\pi$-N scattering to $\pi$-nucleus
scattering~\cite{wein1,silas1}. The non-perturbative effects
responsible for nuclear binding are accounted for using
phenomenological nuclear wavefunctions, as noted above. One can
compute matrix elements using a variety of wavefunctions in order to
ascertain the theoretical error induced by the off-shell behavior of
different wavefunctions.

\begin{figure}[t]
   \vspace{0.5cm} \epsfysize=3.5cm
   \centerline{\epsffile{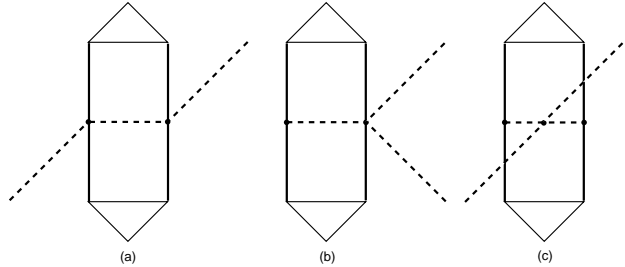}}
   \centerline{\parbox{11cm}{\caption{\label{fig2} 
   Two loop contributions (with an exchanged pion) to the $\pi$-deuteron
   scattering length at $Q^3$.}}}
\end{figure}

To $O({Q^3})$ in $\cpt$ the $\pi$-deuteron scattering length can be written
as~\cite{wein1}

\begin{equation}
a_{\pi d}=\frac{(1+\mu)}{(1+\mu /2)}(a_{\pi n} + a_{\pi p})+{a^{(2a)}}+
{a^{(2b,2c)}},
\label{eq:adfirst}
\end{equation}
where $\mu\equiv{m_\pi}/M$ is the ratio of the pion and the nucleon
mass and the $\pi$-N scattering lengths have the decomposition

\begin{equation}
a_{\pi n} + a_{\pi p}=2 a^{+}={a^{(1a)}}+{a^{(1b)}}
\end{equation}
where $a^+$ is the isoscalar S-wave scattering length and the
superscripts refer to the figures. The various diagrammatic
contributions to $a_{\pi d}$ are illustrated in Fig.~\ref{fig1} and
Fig.~\ref{fig2}. The leading contribution, Fig.~\ref{fig1}(a), has
three nucleon propagators (${Q^{-3}}$), a vertex from ${\cal L}_{\pi
N}^{(2)}$ (${Q^{2}}$), one loop (${Q^{4}}$) and two deuteron
wavefunctions, (${Q^{-1}}$) giving a total of
${Q^{-3+2+4-1}}={Q^{2}}$.  Fig.~\ref{fig1}(b) has five nucleon
propagators (${Q^{-5}}$), one pion propagator (${Q^{-2}}$), three
vertices from ${\cal L}_{\pi N}^{(1)}$ (${Q^{3}}$), two loops
(${Q^{8}}$) and two deuteron wavenfunctions, (${Q^{-1}}$) giving a
total of ${Q^{-5-2+3+8-1}}={Q^{3}}$. One can further verify that the
graphs of Fig.~\ref{fig2} are $O({Q^3})$. Together, Figs.~\ref{fig1}
and \ref{fig2} are all that contribute at $O({Q^3})$ at threshold.

The contributions to $a_{\pi d}$ from the graphs of 
Fig.~\ref{fig2} are:

\begin{equation}
{a^{(2a)}}= - \frac{{m_\pi^2}}{32{\pi^4}{f_\pi^4}{(1+\mu /2)}}
\langle\frac{1}{{\vec q}^{\,2}}\rangle_{\sl wf}
\label{eq:numone}
\end{equation}
\begin{equation}
{a^{(2b,2c)}}=\frac{{g_A^2}{m_\pi^2}}
{128{\pi^4}{f_\pi^4}{(1+\mu /2)}}
\langle\frac{{\vec q}\cdot{{\vec\sigma}_1}{\vec q}\cdot{{\vec\sigma}_2}}
{({\vec q}^{\,2}+{m_\pi^2})^2}\rangle_{\sl wf}
\label{eq:numtwo}
\end{equation}
where $\langle\vartheta\rangle_{\sl wf}$ indicates that $\vartheta$ is
sandwiched between deuteron wavefunctions.  These matrix elements have
been evaluated using a cornucopia of wavefunctions; results are
displayed in table~\ref{tab:exp}. Clearly ${a^{(2a)}}$
dominates. This is the result of the shorter range nature of
$a^{(2b,2c)}$.  It is important to stress that the dominant
contribution from these graphs is quite independent of the
wavefunction used. This implies that the $\cpt$ approach, which relies
on the dominance of the pion-exchange, is useful in this context.

To $O({Q^3})$ in $\cpt$~\cite{bkm1}:

\begin{equation}
4\pi(1+\mu )a^{+}=
\frac{m_\pi^2}{f_\pi^2}\biggl(-4c_1 +2c_2 -\frac{g_A^2}{4M}+2c_3 \biggr)
+\frac{3{g_A^2}{m_\pi^3}}{64\pi{f_\pi^4}},
\end{equation}
where the ${c_i}$ are low-energy constants from ${\cal L}_{\pi
N}^{(2)}$\footnote{ It should be stressed that to this order there
appear large cancellations between the individual terms~\cite{bkm1}
which lead one to suspect that a calculation at $O({Q^4})$ should be
performed to obtain a more precise prediction for this anomalously
small observable.}. The sole undetermined parameter entering the
$O({Q^3})$ computation of $a_{\pi d}$ is therefore a combination of
$c_1$, $c_2$ and $c_3$:

\begin{equation}
\Delta\equiv {-4c_1 +2(c_2 +c_3)}.
\end{equation}

There is recent experimental information about both the $\pi$-N and
$\pi$-deuteron scattering lengths~\cite{chat,sigg}.  Since $a^+$
involves constants that are not fixed by chiral symmetry we can use
experimental information about $\pi$-deuteron scattering to predict $a^+$;
the recent Neuchatel-PSI-ETHZ (NPE) pionic deuterium
measurement~\cite{chat} gives

\begin{equation}
a_{\pi d}=-0.0259 \pm 0.0011\,{m_\pi^{-1}}.
\label{eq:latestpid}
\end{equation}
For the contributions of Fig.~\ref{fig2} we take the average of the
$a^{(2a)}$ and $a^{(2b,2c)}$ values in table~\ref{tab:exp}:

\begin{equation}
a^{(2a+2b+2c)}=-0.0203 \,{m_\pi^{-1}}.
\end{equation}
We then find from Eq.~(\ref{eq:adfirst}):

\begin{equation}
a^{+}=-(2.6 \pm 0.5)\,\cdot\,{10^{-3}}{m_\pi^{-1}}.
\label{eq:apluspred}
\end{equation}
Note that although $a^{(2b,2c)}$ is small, there is a strong
cancellation between $a^{(2a)}$ and $a_{\pi d}$ which leads to a
sensitivity to $a^{(2b,2c)}$.  Our value for $a^+$ is not consistent
with the Karlsruhe-Helsinki value~\cite{koch},

\begin{equation}
a^{+}=-(8.3 \pm 3.8)\,\cdot\,{10^{-3}}{m_\pi^{-1}},
\end{equation}
or the new NPE value deduced from the strong interaction shifts in
pionic hydrogen and deuterium, which is small and 
positive~\cite{sigg}

\begin{equation}
a^{+}=(0...5)\,\cdot\,{10^{-3}}{m_\pi^{-1}}.
\end{equation}
The result Eq.~(\ref{eq:apluspred}) agrees, however, with the value
obtained in the SM95 partial-wave analysis~\cite{vpi}, $a^{+}=-3.0\,\cdot\,
{10^{-3}}{m_\pi^{-1}}$.

\begin{table}[t]
\caption{Contributions of Fig.~\ref{fig2} for various deuteron
wavefunctions in units of $m_\pi^{-1}$.  We use ${f_\pi}=92.4$\,MeV,
${g_A}=1.32$ and ${m_{\pi^+}}=139.6\,$MeV.
\label{tab:exp}}
\vspace{0.2cm}
\begin{center}
\footnotesize
\begin{tabular}{|l|r|r|r|c|}
    \hline ${\sl wf}$ & $a^{(2a)}$ & $a^{(2b,2c)}$ \\ \hline 
    Bonn\protect{~\cite{bonn}} &
  $-0.02021$ & $-0.0005754$ \\ 
   ANL-V18\protect{~\cite{v18}} & $-0.01960$ & $-0.0007919$ \\
   Reid-SC\protect{~\cite{reid}} & 
  $-0.01941$ & $-0.0008499$ \\ 
   SSC\protect{~\cite{ssc}} & 
  $-0.01920$ & $-0.0006987$ \\ \hline 
\end{tabular}
\end{center}
\end{table}

Given the ambiguous experimental situation regarding $a^{+}$, it seems
most profitable to turn our formula around and use the $\pi$-deuteron
scattering data to constrain $\Delta$. We can write

\begin{equation}
\Delta =
\frac{2\pi{f_\pi^2}}{m_\pi^2}(1+\mu /2)
\lbrace a_{\pi d}-({a^{(2a)}}+{a^{(2b,2c)})}\rbrace
+\frac{g_A^2}{4M}\bigl(1-\frac{3M{m_\pi}}{16\pi{f_\pi^2}}\bigr).
\end{equation}
Using Eqs.~(\ref{eq:numone}), (\ref{eq:numtwo}) and (\ref{eq:latestpid})
we find

\begin{equation}
\Delta =-(0.08\pm 0.02)\, {\rm GeV}^{-1},
\end{equation}
where we have taken into account the error in the determination of
$a_{\pi d}$.

In table~\ref{tab:exp2} we give values of the relevant $c_i$
obtained from a realistic fit to low-energy pion-nucleon scattering
data and subthreshold parameters~\cite{bkm2}.  Central values lead to
$\sigma (0)=47.6\,$MeV and $a^+ =-4.7\cdot 10^{-3}{m_\pi^{-1}}$. These
values of the $c_i$ give the conservative determination:

\begin{equation}
\Delta =-(0.18\pm 0.75)\, {\rm GeV}^{-1}.
\end{equation}
Also shown in table~\ref{tab:exp2} are values of the $c_i$ deduced from
resonance saturation. It is worth mentioning that an independent fit
to pion-nucleon scattering including also low-energy constants
related to dimension three operators finds results consistent with the
fit values of table~\ref{tab:exp2}~\cite{moj}.

\begin{table}[t]
\caption{Values of the LECs $c_i$ in GeV$^{-1}$ for $i=1,\ldots,3$.  
Also given are the central values (cv) and the ranges for the $c_i$ from
resonance exchange. The $^*$ denotes an input quantity. 
\label{tab:exp2}}
\vspace{0.2cm}
\begin{center}
\footnotesize
\begin{tabular}{|l|r|r|r|c|}
    \hline
    $i$         & $c_i \quad \quad$   &  
                  $ c_i^{\rm Res} \,\,$ cv & 
                  $ c_i^{\rm Res} \,\,$ ranges    \\
    \hline
    1  &  $-0.93 \pm 0.10$  &  $-0.9^*$ & -- \\
    2  &  $3.34  \pm 0.20$  &  $3.9\,\,$ & $2 \ldots 4$ \\    
    3  &  $-5.29 \pm 0.25$  &  $-5.3\,\,$ 
                                     & $-4.5 \ldots -5.3$ \\
    \hline
    $\Delta$ & $-0.18 \pm 0.75$ & $0.8 \,\, $& $-3.0 \ldots +2.6$\\ 
    \hline
  \end{tabular}
\end{center}
\end{table}

To summarize, we have shown that recent precise data on the
$\pi$-deuteron scattering length can be used to constrain a
combination of dimension two low-energy constants of the pion-nucleon
chiral Lagrangian. This constraint can be improved by going to
$O({Q^4})$ in the chiral expansion. Thus this simple calculation
provides an example of how using effective field theory one can
extract nucleon properties from a nuclear process in a systematic way.

\section{Entr\'ee: $\gamma$-deuteron scattering}
\label{sec:entre}

\subsection{Motivation}\label{subsec:mot}

\noindent Nucleon Compton scattering has been studied in $\cpt$ in
Ref.~\cite{ulf1}, where the following results for the polarizabilities
were obtained to order $Q^3$:

\begin{eqnarray}
\alpha_p=\alpha_n=\frac{5 e^2 g_A^2}{384 \pi^2 f_\pi^2 m_\pi} &=&12.2 \times
10^{-4} \, {\rm fm}^3; \label{eq:alphaOQ3}\nonumber\\ 
\beta_p=\beta_n=\frac{e^2 g_A^2}{768 \pi^2 f_\pi^2 m_\pi}&=& 1.2 \times 
10^{-4} \, {\rm fm}^3. \label{eq:betaOQ3}
\end{eqnarray}
Here we have used $g_A=1.26$ for the axial coupling of the nucleon,
and $f_\pi=93$ MeV as the pion decay constant. Note that the
polarizabilities are {\it predictions} of $\cpt$ at this order.  The
$O(Q^3)$ $\cpt$ predictions diverge in the chiral limit because they
arise from pion loop effects. In less precise language, the
power-counting of $\cpt$ implies that polarizabilities are dominated
by the dynamics of the long-ranged pion cloud surrounding the nucleon,
rather than by short-range dynamics.  The polarizabilities should thus
provide a sensitive test of chiral dynamics.  At the next order in the
chiral expansion, $Q^4$, there are contributions to the
polarizabilities from undetermined parameters which must be fixed
independently~\cite{ulf2}.  These counterterms account for short-range
contributions to the nucleon structure. 

Recent experimental values for the proton polarizabilities are
\cite{newanalysis} \footnote{These are the result of a model-dependent
  fit to data from Compton scattering on the proton at several angles
  and at energies ranging from 33 to 309 MeV.}

\begin{eqnarray}
\alpha_p + \beta_p=13.23 \pm 0.86^{+0.20}_{-0.49} \times 10^{-4} \, {\rm fm}^3,
\nonumber\\
\alpha_p - \beta_p=10.11 \pm 1.74^{+1.22}_{-0.86} \times 10^{-4} \, {\rm fm}^3,
\label{polexpt}
\end{eqnarray}
where the first error is a combined statistical and systematic error,
and the second set of errors comes from the theoretical model
employed. These values are in good agreement with the $\cpt$
predictions.

On the other hand, the neutron polarizabilities are difficult to
obtain experimentally due to the absence of suitable neutron targets
and so the corresponding $\cpt$ prediction is not well tested.  One
way to extract neutron polarizabilities is to consider Compton
scattering on nuclear targets.  Consider coherent photon scattering on
the deuteron. The cross section in the forward direction naively goes
as:

\begin{equation}
\left.\frac{d \sigma}{d \Omega} \right|_{\theta=0}
\sim (f_{Th} - (\alpha_p + \alpha_n) \omega^2)^2.
\end{equation} 
The sum $\alpha_p + \alpha_n$ may then be accessible via its
interference with the dominant Thomson term for the proton,
$f_{Th}$~\cite{hornidge}. This means that with experimental knowledge
of the proton polarizabilities it may be possible to extract those for
the neutron.  Coherent Compton scattering on a deuteron target has
been measured at $E_\gamma=$ 49 and 69 MeV by the Illinois group
\cite{lucas}.  An experiment with tagged photons in the energy range
$E_\gamma= 84.2-104.5$ MeV is under analysis at Saskatoon \cite{SAL},
while data for $E_\gamma$ of about 60 MeV is being analyzed at
Lund~\cite{lund}.

Clearly the amplitude for Compton scattering on the deuteron involves
mechanisms other than Compton scattering on the individual constituent
nucleons. Hence, extraction of nucleon polarizabilities requires a
theoretical calculation of Compton scattering on the deuteron that is
under control in the sense that it accounts for {\it all} mechanisms
to a given order in a systematic expansion in a small parameter.
There exist a few calculations of this reaction in the framework of
conventional potential models \cite{wilbois,levchuk,jerry}.  These
calculations yield similar results if similar input is supplied, but
typically mechanisms for nucleon polarizabilities and two-nucleon
contributions are not treated consistently.  We will see that $\cpt$
provides an alternative framework where this drawback can be
eliminated.

In the remainder of this paper I will review a recent computation of
Compton scattering on the deuteron for incoming photon energies of
order 100 MeV in the Weinberg formulation. As in the computation of
the pion-deuteron scattering length, baryon $\cpt$ is used to compute
an irreducible scattering kernel to order $Q^3$, which is then sewn to
external deuteron wavefunctions.

\begin{figure}[t]
   \vspace{0.5cm} \epsfysize=3.5cm
   \centerline{\epsffile{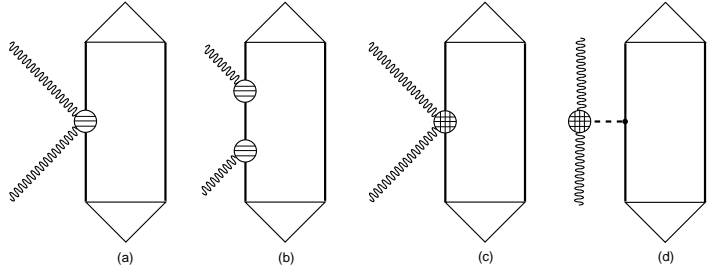}}
   \centerline{\parbox{11cm}{\caption{\label{fig5} One loop graphs
   which contribute to Compton scattering on the deuteron at order
   $Q^2$ (a) and at order $Q^3$ (b-d) (in the Coulomb gauge).  The
   sliced and diced blobs are from ${\cal L}_{\pi N}^{(3)}$ (c) and
   ${\cal L}_{\pi \gamma}^{(4)}$ (d).  Crossed graphs are not
   shown.}}}
\end{figure}

\begin{figure}[t]
   \vspace{0.5cm} \epsfysize=6.5cm
   \centerline{\epsffile{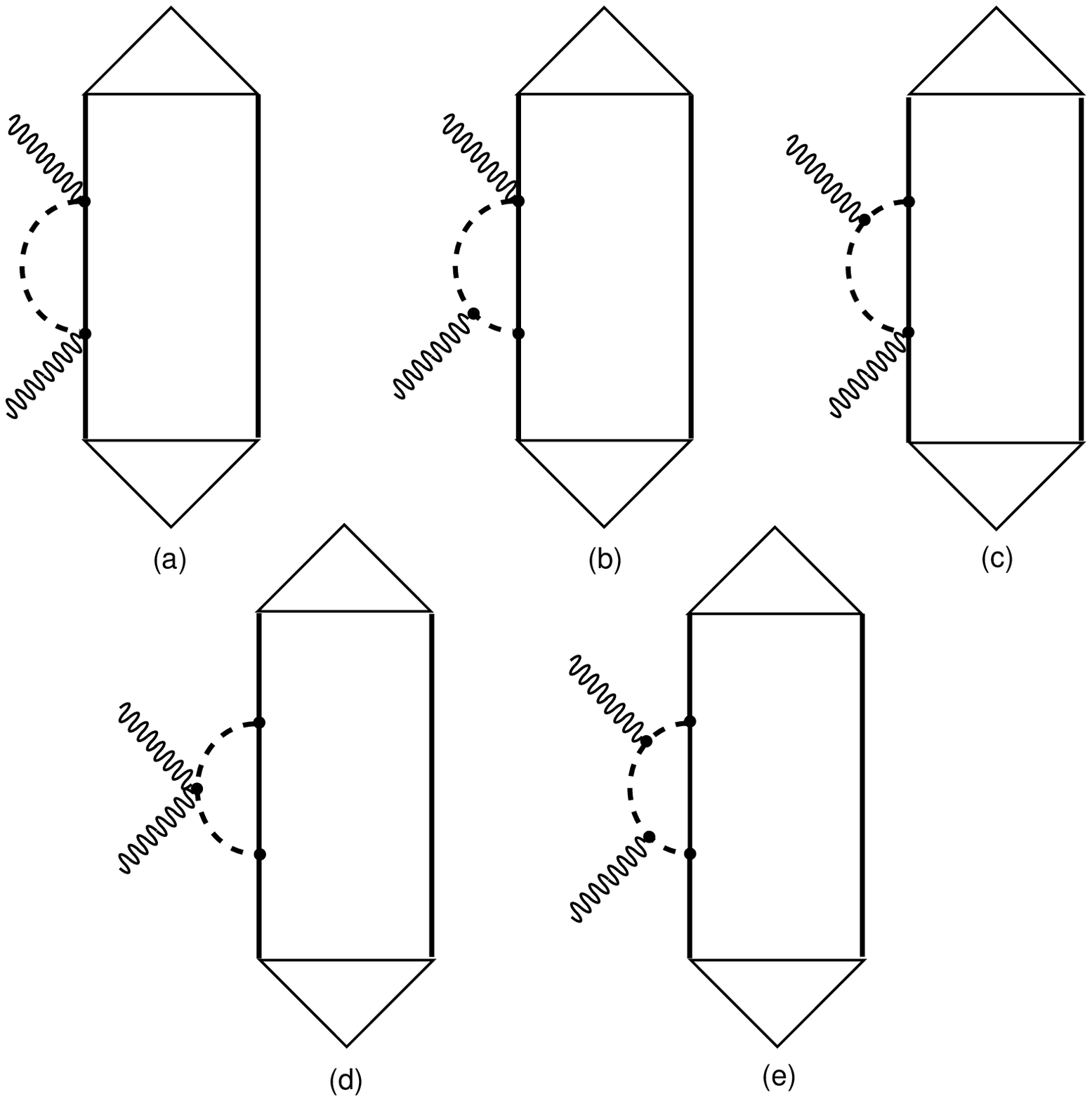}}
   \centerline{\parbox{11cm}{\caption{\label{fig6} Two loop graphs
   which contribute to Compton scattering on the deuteron at order
   $Q^3$ Crossed graphs are not shown.}}}
\end{figure}

\begin{figure}[t,h,b,p]
   \vspace{0.5cm}
   \epsfysize=6.5cm
   \centerline{\epsffile{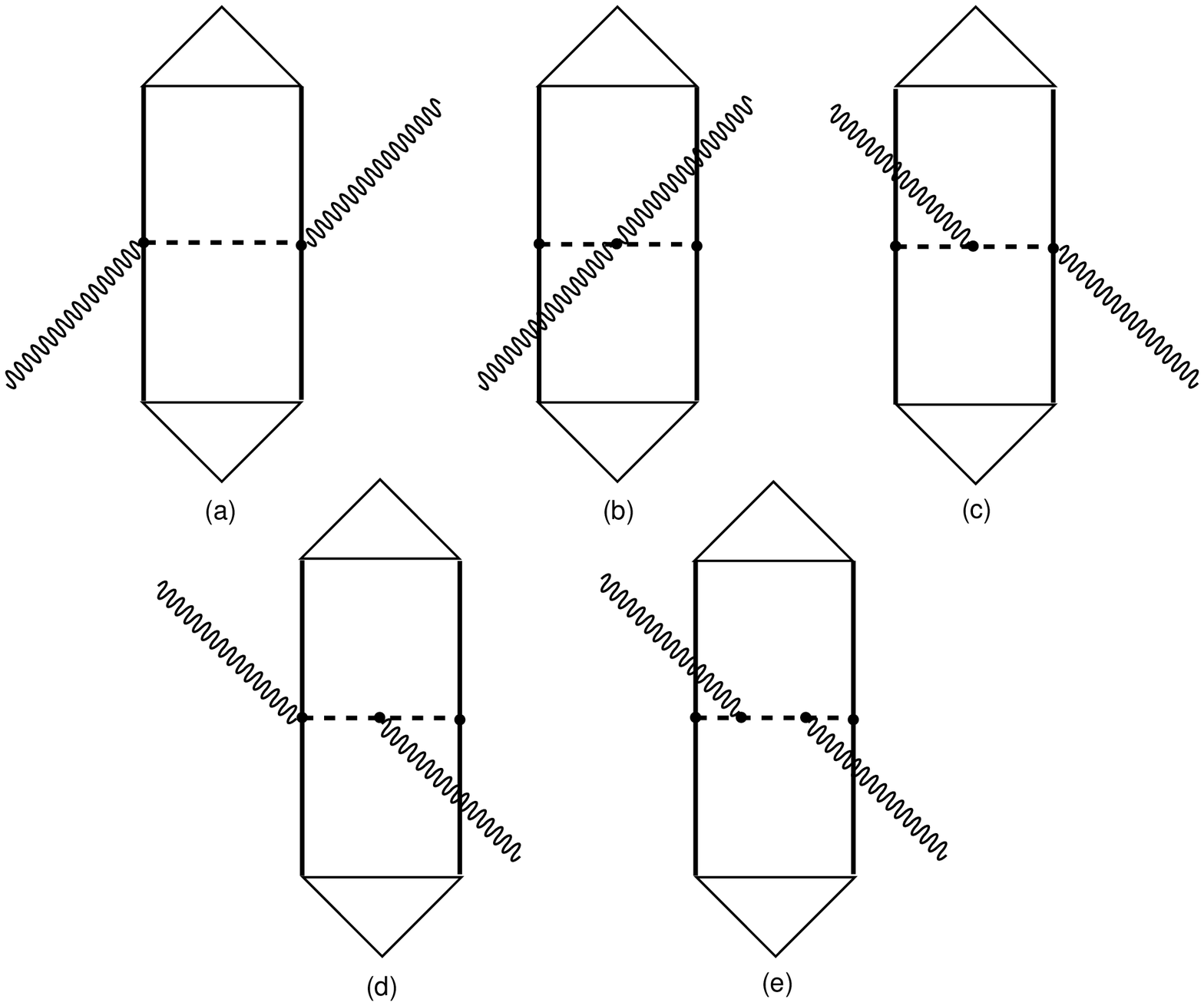}}
   \centerline{\parbox{11cm}{\caption{\label{fig7}
   Two loop graphs which contribute to Compton scattering
   on the deuteron at order $Q^3$. Crossed graphs are not shown. }}}
\end{figure}

\subsection{Compton scattering to $O(Q^3)$}\label{subsec:comptcalc}

\noindent The Compton amplitude we wish to evaluate is (in the $\gamma d$
center-of-mass frame):

\begin{eqnarray}
T^{\gamma d}_{M' \lambda' M \lambda}(\vkayprime,\vkay)&=& \int
\frac{d^3p}{(2 \pi)^3} \, \, \psi_{M'}\left( \vpee + \frac{\vkay -
\vkayprime}{2}\right) \, \, T^{\gamma d \, \, c.m.}_{\gamma N_{\lambda'
\lambda}}(\vkayprime,\vkay) \, \, \psi_M(\vpee)\nonumber\\ 
&+& \int \frac{d^3p \, \, d^3p'}{(2 \pi)^6} \, \, \psi_{M'}(\vpeeprime) \, \,
T^{2N}_{\gamma NN_{\lambda' \lambda}}(\vkayprime,\vkay) \, \, \psi_M(\vpee)
\label{eq:gammad}
\end{eqnarray}
where $M$ ($M'$) is the initial (final) deuteron spin state, and
$\lambda$ ($\lambda'$) is the initial (final) photon polarization
state, and $\vkay$ ($\vkayprime$) the initial (final) photon
three-momentum, which are constrained to
$|\vkay|=|\vkayprime|=\omega$.  The amplitude $T^{\gamma d \, \,
c.m.}_{\gamma N}$ represents the graphs of figures~\ref{fig5} and
\ref{fig6} where the photon interacts with only one nucleon.  Of
course this amplitude must be evaluated in the $\gamma d$
center-of-mass frame. The amplitude $T^{2N}_{\gamma NN}$ represents
the graphs of Fig.~\ref{fig7} where there is an exchanged pion
between the two nucleons.

The leading contribution to Compton scattering on the deuteron is
shown in Fig.~\ref{fig5}(a). This graph has three nucleon
propagators (${Q^{-3}}$), a vertex from the second order pion-nucleon
Lagrangian (${Q^{2}}$), one loop (${Q^{4}}$) and two deuteron
wavefunctions, (${Q^{-1}}$) giving a total of
${Q^{-3+2+4-1}}={Q^{2}}$. This contribution is simply the Thomson term
for scattering on the proton.  At next order, $O({Q^{3}})$, there are
several more one loop graphs Fig.~\ref{fig5}(b,c,d) and two loop
graphs without (Fig.~\ref{fig6}) and with (Fig.~\ref{fig7}) a pion
exchanged between the nucleons.  The full amplitudes are given in
Ref.~\cite{silas3}.

For the wave function $\psi$ we use the energy-independent Bonn OBEPQ
wave function parameterization which is found in Ref.~\cite{Bonnrep}.
The photon-deuteron $T$-matrix (\ref{eq:gammad}) is then calculated
and the laboratory differential cross section evaluated directly from
it:

\begin{equation}
  \frac{d \sigma}{d \Omega_L}=\frac{1}{16 \pi^2}
  \left(\frac{E_\gamma'}{E_\gamma}\right)^2 \frac{1}{6} \sum_{M'
    \lambda' M \lambda} |T^{\gamma d}_{M' \lambda' M \lambda}|^2,
\end{equation}
where $E_\gamma$ is the initial photon energy in the laboratory frame
and $E_\gamma'$ is the final photon energy in the laboratory frame.

Convergence tests indicate that with the numbers of quadratures chosen
the cross section evaluated in this fashion is numerically accurate at
about the 1\% level. Of course, this error does not include the
theoretical error from uncertainties due to different deuteron wave
functions\footnote{As discussed in detail in Ref.~\cite{silas3},
wavefunction errors are minimal and well understood.}, and the effect
of higher-order terms in $\cpt$.  The errors due to omitted higher
orders in $\cpt$ will be discussed further below.

\begin{figure}[t,h,b,p]
   \vspace{0.5cm} \epsfysize=6cm
   \centerline{\epsffile{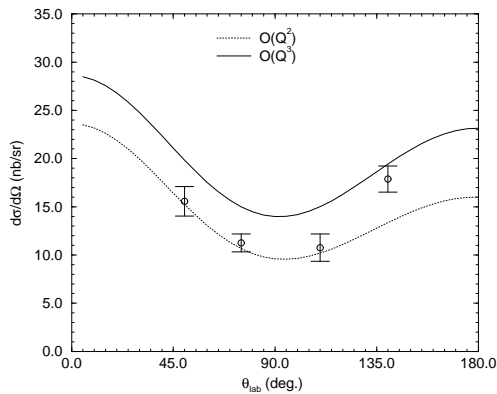}}
   \centerline{\parbox{11cm}{\caption{\label{fig8} Results of the
   $O(Q^2)$ (dotted line) and $O(Q^3)$ (solid line) calculations
   at a photon laboratory energy of 49 MeV. }}}
\end{figure}

\begin{figure}[t,h,b,p]
   \vspace{0.5cm} \epsfysize=6cm
   \centerline{\epsffile{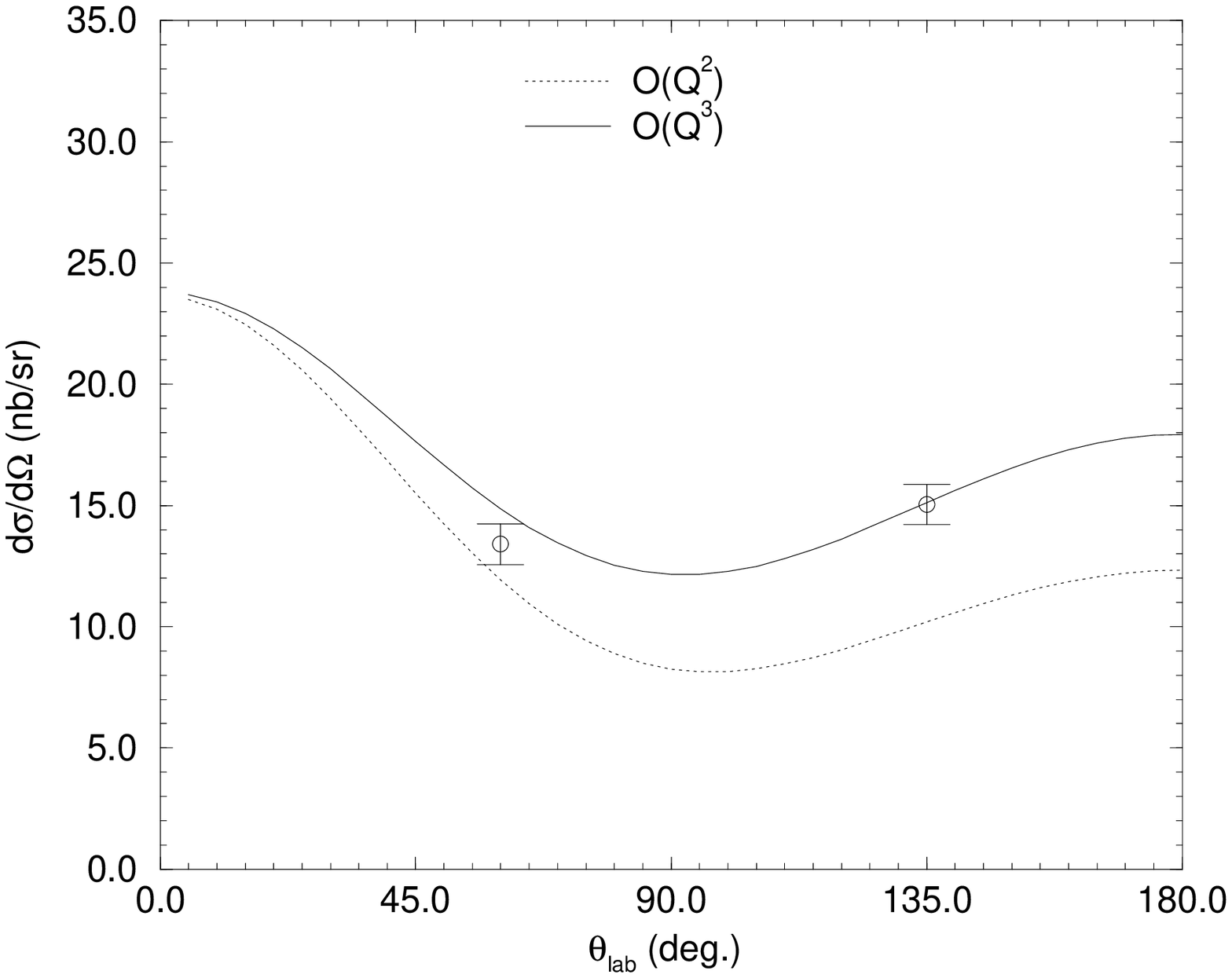}}
   \centerline{\parbox{11cm}{\caption{\label{fig9} Results of the
   $O(Q^2)$ (dotted line) and $O(Q^3)$ (solid line) calculations
   at a photon laboratory energy of 69 MeV.}}}
\end{figure}

\begin{figure}[t,h,b,p]
   \vspace{0.5cm} \epsfysize=6cm
   \centerline{\epsffile{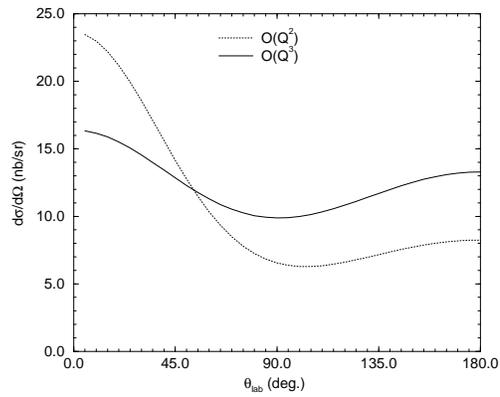}}
   \centerline{\parbox{11cm}{\caption{\label{fig10} Results of the
   $O(Q^2)$ (dotted line) and $O(Q^3)$ (solid line) calculations
   at a photon laboratory energy of 95 MeV.}}}
\end{figure}

In figures~\ref{fig8}, \ref{fig9} and \ref{fig10} we display our
results at 49, 69, and 95 MeV. For comparison we have included the
calculation at $O(Q^2)$, where the second contribution in
Eq.~(\ref{eq:gammad}) is zero, and the $\gamma N$ $T$-matrix in the
single-scattering contribution is given by the Thomson term on a
single nucleon. It is remarkable that to $O(Q^3)$ no unknown
counterterms appear.  All contributions to the kernel are fixed in
terms of known pion and nucleon parameters such as $m_\pi$, $g_A$,
$M$, and $f_\pi$.  Thus, to this order $\cpt$ makes {\it predictions}
for Compton scattering.

The curves show that the correction from the $O(Q^3)$ terms gets
larger as $\omega$ is increased, as was to be expected. Indeed, while
at lower energies corrections are relatively small, in the 95 MeV
results the correction to the differential cross section from the
$O(Q^3)$ terms is of order 50\%, although the contribution of these
terms to the {\it amplitude} is of roughly the size one would expect
from the power-counting: about 25\%.  Nevertheless, it is clear, even
from these results, that this calculation must be performed to
$O(Q^4)$ before conclusions can be drawn about polarizabilities from
data at photon energies of order $m_\pi$. This is in accord with
similar convergence properties for the analogous calculation for
threshold pion photoproduction on the deuteron~\cite{silas2}.

We have also shown the six Illinois data points at 49 and 69
MeV~\cite{lucas}. Statistical and systematic errors have been added in
quadrature. It is quite remarkable how well the $O(Q^2)$ calculation
reproduces the 49 MeV data. However, it is clear that the agreement at
forward angles is somewhat fortuitous, as there are significant
$O(Q^3)$ corrections.  Meanwhile, the agreement of the $O(Q^3)$
calculation with the 69 MeV data is very good, although only limited
conclusions can be drawn, given that there are only two data points,
each with sizeable error bars.

\begin{figure}[t,h,b,p]
   \vspace{0.5cm} \epsfysize=6cm
   \centerline{\epsffile{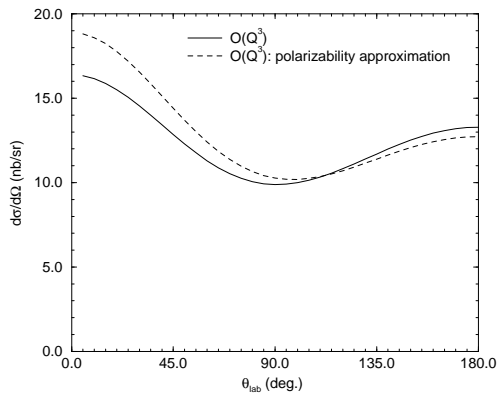}}
   \centerline{\parbox{11cm}{\caption{\label{fig11}
   Unpolarized cross section for Compton scattering on the deuteron in
   deltaless $\cpt$ to $O(Q^3)$ at 95 MeV.  The polarizability
   approximation to the single-nucleon amplitude (dashed line) is to
   be compared with the full $O(Q^3)$ calculation (solid line).}}}
\end{figure}

Our results are qualitatively not very different from other existing
calculations.  At 49 and 69 MeV our $O(Q^3)$ results are very close to
those in Ref. \cite{wilbois} and a few nb/sr higher, especially at
back angles, than those of Refs. \cite{levchuk,jerry} (which are
similar at these energies).  At 95 MeV our $O(Q^3)$ result is close to
that of Ref. \cite{levchuk}, higher by several nb/sr at back angles
than Ref. \cite{jerry}, and several nb/sr lower than the calculation
with no polarizabilities of Ref. \cite{wilbois}~\footnote{At this
energy Ref. \cite{wilbois} only presents results with
$\alpha_p+\alpha_n=\beta_p+\beta_n=0$, which in turn are considerably
less forward peaked than the corresponding calculation of
Ref. \cite{levchuk}.}.  Comparing to the calculations of deuteron
Compton scattering in the KSW formulation of effective field
theory~\cite{martinetal}, we see that the result of Ref.
\cite{martinetal} is significantly lower than those presented here at
both 49 and 69 MeV. At 49 MeV the agreement of
Ref.~\cite{martinetal}'s calculation with the data is better than
ours. We shall show in the next section that this is partly because 49
MeV is at the lower end of the domain of applicability of the Weinberg
formulation.  At 69 MeV our calculation does a slightly better job of
reproducing the (two) data points available. The qualitative agreement
among these calculations is a reflection of the similarities of
mechanisms involved.  Ours is however the only calculation to
incorporate the full single-nucleon amplitude instead of its
polarizability approximation.  As shown in Fig.~\ref{fig11} our
tendency to higher relative cross sections in the backward directions
is at least in part due to this feature.

\subsection{Effective theories of Compton scattering}\label{subsec:effth}

\begin{figure}[t,h,b,p]
   \vspace{0.5cm} \epsfysize=3.5cm
   \centerline{\epsffile{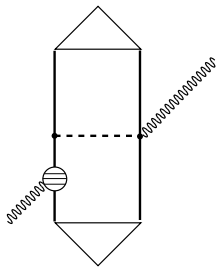}}
   \centerline{\parbox{11cm}{\caption{\label{fig3} Interaction which
   contributes to Compton scattering on the deuteron at order
   $Q^4$. The sliced blob represents a $1/M$ correction vertex from
   ${\cal L}_{\pi N}^{(2)}$.}}}
\end{figure}

\noindent Although nominally the domain of validity of the Weinberg
formulation extends well beyond the threshold for pion production, the
power-counting fails at low energies well before the Thompson limit is
reached.  Consider the $O(Q^4)$ contribution shown in Fig. \ref{fig3}.
We can use this graph to illustrate the transition to the very-low
energy regime $Q\sim m_\pi^2/M$.  It is easy to see that this graph
becomes comparable to the order $Q^3$ graph of Fig. \ref{fig7}(a) when

\begin{equation}
{\frac{|\vpee\; |^2}{\omega M}}\sim 1.
\end{equation}
Here $\vpee$ is a typical nucleon momentum inside the deuteron and
$\omega$ is the photon energy.  Since our power-counting is predicated
on the assumption that all momenta are of order $m_\pi$, we find that
power-counting is valid in the region

\begin{equation}
\frac{m_\pi^2}{M}\ll Q \ll \lsc .
\end{equation}
Therefore, in the region $\omega \sim B$ the Weinberg power-counting
is not valid, since the external probe momentum flowing through the
nucleon lines is of order $Q^2/M$, rather than order $Q$. It is in
this region that the Compton low-energy theorems are
derived. Therefore our power-counting will not recover those
low-energy theorems. Of course the upper bound on the validity of the
effective theory should increase if the $\Delta$-resonance is included
as a fundamental degree of freedom~\cite{jenkinsetal}.

In Ref.~\cite{martinetal} Compton scattering on the deuteron
was computed to the same order discussed here, one order beyond
leading non-vanishing order but in the KSW formulation of two-nucleon
effective field theory. An advantage of KSW power-counting is that the
effective field theory moves smoothly between $Q < B$ and $Q > B$.
KSW power-counting is valid for nucleon momenta $Q< {\Lambda_{NN}}\sim
300$ MeV. Thus in the KSW formulation deuteron polarizabilities and
Compton scattering up to energies $\omega < {\Lambda_{NN}^2}/{M}\sim
90$ MeV can be discussed in the same framework.  Here we are
interested mostly in the region $\omega\sim m_\pi$, and so we regard
ourselves as being firmly in the second regime.  We
stress that the value of ${\Lambda_{NN}}$ is uncertain; it is
conceivable that ${\Lambda_{NN}}\sim 500$ MeV in which case the range
of the KSW formulation would extend well beyond pion production
threshold.

\subsection{Effects of higher order terms}\label{subsec:higher}

In order to test the sensitivity of our calculation to higher-order
effects we added a small piece of the $O(Q^4)$ amplitude for Compton
scattering off a single nucleon.  Specifically, we have modified the
Compton amplitudes so that the polarizabilities in the calculation are
changed to

\begin{equation}
\alpha=\alpha^{(Q^3)} + \Delta \alpha, \qquad \qquad
\beta=\beta^{(Q^3)} + \Delta \beta,
\label{mockpol}
\end{equation}
where $\alpha^{(Q^3)}$ and $\beta^{(Q^3)}$ are the $O(Q^3)$ values of
Eq.~(\ref{eq:betaOQ3}).  We emphasize that this effect represents
only a part of what will appear in the single-nucleon scattering
amplitude $T_{\gamma N}$ at $O(Q^4)$. Furthermore, a number of
additional mechanisms must be included in $T_{\gamma NN}^{2N}$ in any
$O(Q^4)$ calculation of Compton scattering on the
deuteron. Nevertheless, here we calculate the differential cross
section with the modified polarizabilities in order to get a feel for
the sensitivity of our result to the presence of such higher-order
terms.

\begin{figure}[t,h,b,p]
   \vspace{0.5cm} \epsfysize=6cm
   \centerline{\epsffile{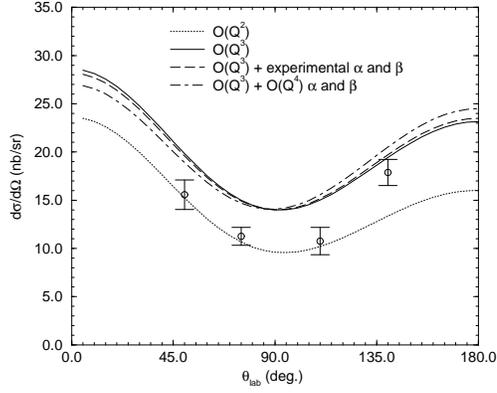}}
   \centerline{\parbox{11cm}{\caption{\label{fig12} Results
   of calculations at 49 MeV using different values for the
   nucleon electromagnetic polarizabilities. The solid line is
   the result using the $O(Q^3)$ $\cpt$ value, the long-dashed
   line is the result using ``experimental'' polarizabilities, and
   the dot-dashed line represents a calculation with the $O(Q^4)$
   polarizabilities.}}}
\end{figure}

\begin{figure}[t,h,b,p]
   \vspace{0.5cm} \epsfysize=6cm
   \centerline{\epsffile{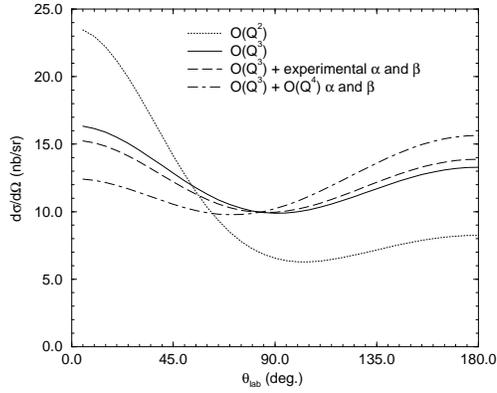}}
   \centerline{\parbox{11cm}{\caption{\label{fig13} Results
   of calculations at 95 MeV using different values for the
   nucleon electromagnetic polarizabilities. Legend as in
   Fig.~\protect{\ref{fig12}}.}}}
\end{figure}

Two calculations were performed.  In the first, $\Delta \alpha$ and
$\Delta \beta$ were chosen so that the total polarizabilities
(\ref{mockpol}) were equal to the ``experimental''
values\footnote{Here we use the proton experimental values and neutron
``experimental'' values (see Ref.~\cite{silas3} for details).}.  The
second calculation involved a more dramatic change in the
polarizabilities: $\Delta \alpha$ and $\Delta \beta$ were chosen so
that $\alpha$ and $\beta$ were equal to the $O(Q^4)$ values of
Ref.~\cite{ulf2} where resonance-saturation has been used to estimate
some of the $O(Q^4)$ $\cpt$ counterterms. In either case $\Delta
\alpha_p+\Delta \alpha_n$ is relatively small, while $\Delta
\beta_p+\Delta \beta_n$ is not large for ``experimental'' values, but
is significant for the $O(Q^4)$ values.  The results of these two
calculations for the two photon energies $E_\gamma=49$ MeV and
$E_\gamma=95$ MeV are shown in Fig.~\ref{fig12} and Fig.~\ref{fig13}.

In both cases we see that, just as one would expect, the cross section
at 95 MeV is much more sensitive to these $O(Q^4)$ terms than the
cross section at 49 MeV.  It is not surprising that the calculation
with $O(Q^4)$ polarizabilities exhibits a larger change than that with
``experimental'' values.  Continually increasing $\beta_p+\beta_n$ at
approximately constant $\alpha_p+\alpha_n$ decreases the cross section
at forward angles and increases it at back angles.  In fact, it seems
that if $\beta_p+\beta_n$ is sufficiently large then the character of
the cross section at 95 MeV can change completely from forward peaked
to backward peaked. 

Significant change in the 95 MeV results presented here from those
obtained at $O(Q^3)$ mandates a cautious interpretation.  At the same
time that the cross section at 95 MeV is more sensitive to
polarizabilities than at lower energies, it is also more sensitive to
$O(Q^4)$ corrections. In our view, a full $O(Q^4)$ calculation in
$\cpt$ is necessary if any attempt is to be made to extract the
neutron polarizability from the Saskatoon data within this framework.

\section{Dessert}
\label{sec:conc}

\noindent Using effective field theory techniques, nucleon properties
can be extracted systematically from experiments performed with
nuclei.  As a simple example, we have reviewed a $\cpt$ calculation
which uses $\pi$-deuteron data to place constraints on low energy
constants of the $\pi$-N chiral Lagrangian.

And, too, we have reviewed a recent computation of Compton scattering
on the deuteron in $\cpt$. We find reasonable agreement with the data
at 49 MeV.  At this energy $O(Q^3)$ corrections are not large compared
to the leading $O(Q^2)$ result, and $O(Q^4)$ terms seem to be even
smaller.  However, as anticipated, the effective theory appears to
break down as the Thomson limit is approached. We find good agreement
with the data at 69 MeV.  At this energy the convergence appears to be
good. This suggests that $\cpt$ at $O(Q^3)$ is providing reasonable
neutron and two-nucleon contributions.  We find that the
polarizability approximation should not be used in the calculation of
the differential cross section at 95 MeV, since truncating the
photon-nucleon amplitude at order $\omega^2$ results in a significant
change in the photon-deuteron differential cross section for forward
angles. The wave function dependence is minor (on the order of 10\% in
the differential cross section).  We make a prediction at 95 MeV which
is, however, plagued by considerable uncertainties. Convergence is
slow at this energy, as indicated by the relative size of both the
full set of $O(Q^3)$ corrections and a partial set of $O(Q^4)$
corrections.  The cross section tends to come out somewhat smaller
than at lower energies, in particular in the backward directions,
although the full $O(Q^4)$ amplitude is likely to be somewhat bigger
at back angles.  It seems that a more stringent test of $\cpt$ at
these energies---including aspects of neutron structure beyond the
$O(Q^3)$ ``pion cloud'' picture---will have to wait for a next-order
calculation.

\section*{Acknowledgments}
I thank Veronique Bernard, Harry Lee, Man\'e Malheiro, Ulf
Mei{\ss}ner, Dan Phillips and Ubi van Kolck for enjoyable
collaborations, Dan and Ubi for comments on the MS and the INT for a
great workshop.  This research was supported by DOE grant
DE-FG02-93ER-40762 (DOE/ER/40762-183, UMPP\#99-112).

\end{document}